# Spin Coherence and Echo Modulation of the Silicon Vacancy in 4H-SiC at Room Temperature

S. G. Carter[1], Ö. O. Soykal[2], Pratibha Dev[2], Sophia E. Economou[1], E. R. Glaser[1]

[1] Naval Research Laboratory, Washington, DC 20375
[2] National Research Council Research Associate at the Naval Research Laboratory, Washington, DC 20375

The silicon vacancy in silicon carbide is a strong emergent candidate for applications in quantum information processing and sensing. We perform room temperature optically-detected magnetic resonance and spin echo measurements on an ensemble of vacancies and find the properties depend strongly on magnetic field. The spin echo decay time varies from less than 10 μs at low fields to 80 μs at 68 mT, and a strong field-dependent spin echo modulation is also observed. The modulation is attributed to the interaction with nuclear spins and is well-described by a theoretical model.

Deep defect centers in solids are of great current interest as quantum bits or quantum emitters for applications in quantum computing, communication, and sensing, as they combine strengths from the solid state and the atomic world. In particular, the electronic and spin states of some defects have many desirable properties including high efficiency emission of single photons [1–4], highly coherent spin states even at room temperature [5–10], and optical initialization and readout [11–13]. Nitrogen-vacancy (NV) centers in diamond, consisting of a nitrogen atom substituted for a carbon next to a vacancy, have been extensively studied and have thus become the standard for such defects.

There is currently a strong interest to investigate similar defects in other material systems that may have improved properties for quantum applications. SiC has particularly attractive features, including significantly lower cost, mature microfabrication [14–19], and emission in the near-infrared where loss is reduced. A number of defects in silicon carbide (SiC) have significant potential including the antisite-vacancy defect [3,20], the Si vacancy [10,13,21–24], and the divacancy [7,9,25–27]. Experiments have shown that some of these defects can have similar coherence and optical manipulation properties as NV centers [7,10,12,27]. The properties of these defects, however, are not as well-known as for the NV center, and there are several types of each defect that vary to some extent in the different polytypes of SiC.

The focus of this Letter is on the Si vacancy in SiC, a defect that has a distinct spin structure from other defects in SiC and NV centers and which as such may offer additional capabilities [28–30]. This defect allows for efficient optical spin initialization and readout [12,24] as well as coherent microwave manipulation [12,31]. Recently these capabilities have been demonstrated for an isolated vacancy, and the first spin echo measurements were performed, giving a decoherence time on the order of 100 μs at 27 and 28.8 mT [10]. However, very little is known of the spin coherence of this defect beyond this measurement, and even the nature of the spin structure and transitions have been a source of confusion [10,21,22,28,29,31]. We perform optically-detected magnetic resonance (ODMR) and spin echo measurements on an ensemble of Si vacancies at room temperature. We carefully map out the ODMR as a function of magnetic field, providing a clearer picture of the spin transitions of this S=3/2 system [28,29,31,32]. We also perform the first magnetic field dependence of the spin coherence that reveals a strong dependence of the echo decay time on magnetic field and a strong echo modulation. These results agree very well with a detailed theoretical model that takes into account the unique spin structure and the hyperfine interaction with nearby nuclear spins.

Experiments are performed on high purity semi-insulating 4H-SiC. To generate Si vacancies, it is irradiated with 2 MeV electrons at a dose of $5\times10^{17}$ cm$^{-2}$. Figure 1(b) displays the photoluminescence at low temperature (19 K) and room temperature. At low temperature there are several sharp lines, two of which at 1438 meV and 1353 meV have previously been identified as the zero-phonon lines (ZPL) of Si vacancies at the two inequivalent sites, labeled V1 and V2 [22]. At room temperature, the emission consists of broadened ZPL and phonon sidebands from Si vacancies and likely emission from other defects.

ODMR measurements are performed at room temperature as illustrated in Fig. 1(a) by measuring changes in the PL when driving the system with a microwave magnetic field. The excitation laser is focused onto the sample with a 0.62 NA 50X objective that also collects PL and sends it to a Si photodiode. A microwave magnetic field is produced by shorting the inner conductor of a coax line to the outer conductor with a 50 μm diameter gold wire. The sample is oriented on its side in-between the poles of an electromagnet with $B_{static}$ parallel to the c-axis.

The ODMR spectrum at 34 mT is displayed in Fig. 1(c) for three microwave powers, showing two strong lines separated by 140 MHz, which correspond to V2. There is no sign of V1 in the spectrum, consistent with other room temperature measurements [10,28,29]. The weaker pair of lines split by 70 MHz will be discussed later. The two strong lines are broadened at the highest microwave power, but at low powers each line is shown to be composed of three lines [see Fig. 1(d)]. The two weaker lines separated by 8 MHz correspond to vacancies with a next-nearest-neighbor (NNN) $^{29}$Si nuclei (I=1/2), which shift the transition due to the hyperfine interaction [23].

Figure 2(c) displays a map of ODMR as a function of both magnetic field and microwave frequency. At non-zero magnetic fields, there are clearly four transitions, with the stronger two having a slope corresponding to $\Delta m_s = \pm 1$ with g=2, and the weaker two having double the slope, corresponding to $\Delta m_s = \pm 2$. The presence of these four transitions with different slopes is a clear indication that this is a S=3/2 system instead of an S=1 system, which could have at most three transitions. The spin Hamiltonian for this system is $H = g\mu_B \vec{B} \cdot \vec{S} + DS_z^2 + H_{hf}$, where $\vec{B}$ is the magnetic field vector, $z$ is along the c-axis, $2D$ is the zero-field splitting, and $H_{hf}$ is the hyperfine interaction. The four calculated energy levels for S=3/2 are plotted as a function of $B_z$ in Fig. 2(a). Only $\Delta m_s = \pm 1$ transitions are magnetic dipole allowed, suggesting that only $-3/2 \rightarrow -1/2$, $+1/2 \rightarrow +3/2$, and $-1/2 \rightarrow +1/2$ should occur (labeled *a*, *b*, and *c*, respectively), but mixing due to a stray B-field perpendicular to the c-axis or the hyperfine interaction allows weak $\Delta m_s = \pm 2$ transitions, $-3/2 \rightarrow +1/2$ and $-1/2 \rightarrow +3/2$ (labeled *d* and *e*). While transition *c* is allowed, it does not appear in the spectrum because of the mechanism for ODMR, in which spin-dependent processes are only selective between the $m_s = \pm 1/2$ and $\pm 3/2$ states [24,28,33].

Figure 2(d) displays model calculations of the ODMR, with good qualitative agreement with experiment [33]. A level anticrosing (LAC) between $m_s = -1/2$ and $-3/2$ is apparent at 2.5 mT, which strongly reduces the ODMR signal. The strength of the LAC is determined by mixing induced by a magnetic field perpendicular to the c-axis. A perpendicular field of 0.05 mT is included, estimating the average field from NNN $^{29}$Si nuclei. In Fig. 2(e) the strength of the ODMR lines *a* and *b* are plotted as a function of magnetic field, illustrating the intensity dip at 2.5 mT and a weaker, broader dip at 16 mT that we assign to a LAC of ES spin levels. At fields far above the GS LAC [as in Fig. 1(c)], only transitions *a* and *b* remain with a splitting of 4D=140 MHz. At high microwave power, two-photon versions of transitions *d* and *e* are also present in Fig. 1(c) [28], with a splitting of 2D = 70 MHz.

We have also performed some of the first measurements of the coherence time of the Si vacancy in SiC using Hahn spin echo sequences to eliminate the effects of inhomogeneous broadening and slow variations in the environment. The sequence in Fig. 3(a) consists of a NIR pulse of 2 μs to polarize and read-out the spins simultaneously, followed by three microwave pulses with lengths designed to give a π/2-π-π/2 sequence. The microwave pulse lengths were determined by measuring Rabi oscillations with a single microwave pulse of variable length, as plotted in Fig. 3(b). The oscillations are heavily damped, due to the relatively short inhomogeneous dephasing time $T_2^*$.

The spin echo measurements at 34 mT in Fig. 3(c,d) are obtained by setting the microwave frequency to transition *b*, and scanning *t*, the third pulse delay relative to the first pulse, for a series of delays T of the π-pulse (2$^{nd}$ pulse). In Fig. 3(c) the spin echo is displayed for T = 1 μs, which appears as a sharp dip at 2 μs with rapidly decaying oscillations at a frequency of 4.1 MHz. The oscillations are consistent with the hyperfine splitting that gives three different transition frequencies beating with each other. The decay time of 190 ns corresponds to $T_2^*$, consistent with the low power transition linewidth.

In Fig. 3(d), the echo amplitude appears to oscillate with T without clear decay. To better study this behavior we display the amplitude of the echo as a function of T for a series of magnetic fields in Fig. 4(a-e), going out to T = 30 μs. There is clearly echo modulation at all these fields. The oscillations do not have a single frequency, but the modulation timescales are comparable to the Zeeman precession periods of $^{13}$C and $^{29}$Si. (The nuclear Zeeman periods are $2\pi/\omega_{Si}$ = 11.8 μs and $2\pi/\omega_C$ = 9.35 μs at 10 mT.) The modulation amplitude also appears to decrease at the highest field. There is little decay of the echo at 68 mT on this time scale, so we also measured the echo amplitude out to T = 55 μs in Fig. 4(f). From an exponential fit, the echo decay time is 81±4 μs, similar to the value obtained in Ref. [10] for a single Si vacancy. Clearly the decay is much faster for fields lower than 20 mT despite weak revivals at longer delays, falling below $1/e$ of the maximum by T = 3-4 μs.

The spin echo data shows Electron Spin Echo Envelope Modulation [34,35] that results from the anisotropic spin-spin interactions between the unpaired electrons in the vacancy and nearby nuclei. We theoretically describe this behavior with the following spin Hamiltonian: $H = g\mu_B B S_z + D S_z^2 + \omega_I I_z + S_z A I_z + S_z A' I_x + \Omega S_x$. The magnetic field is oriented along the c-axis, $\Omega$ is the Rabi frequency from the microwave drive field, and both nuclear species with non-zero spin, $^{29}$Si and $^{13}$C (both I = ½), are considered. The $A$-term in the Hamiltonian contains contributions from the isotropic (Fermi contact interaction) and anisotropic dipole-dipole hyperfine interaction. $A'$ is the pseudosecular contribution from the anisotropic hyperfine interaction. Both $A$ and $A'$ depend on the vacancy-nucleus distance and the orientation of the pair with respect to the c-axis. Due to the $A'$ term, the quantization axis of the nuclear spin is not aligned along $z$ and changes with the spin state of the vacancy. Nuclear spin precession thus results in a time-varying effective magnetic field for the electron spin. This results in imperfect rephasing when there is significant nuclear precession during time T. Even though all four levels of $m_s = \pm 3/2, \pm 1/2$ are included in our calculations, consideration of only the directly driven $m_s$= +1/2 and +3/2 states can be used to obtain a simplified analytical form for the echo signal from a single nuclear site that is proportional to $1 - \frac{2\omega_I^2 A'^2}{\omega_{3/2}^2 \omega_{1/2}^2} \sin^2 \frac{\omega_{3/2}\tau}{2} \sin^2 \frac{\omega_{1/2}\tau}{2}$, where $\omega_{j=1/2,3/2} = \sqrt{(\omega_I + jA)^2 + (jA')^2}$ are the nuclear precession frequencies when the electronic system is in the $m_s$= +1/2 or +3/2 state. This mechanism for echo modulation appears to have essentially the same origin as that observed for NV centers in diamond [36,37], with the exception that there is no $m_s$ = 0 state in which the hyperfine interaction is absent. The modulation frequency is always sensitive to the strength of the hyperfine interaction here. In turn, the hyperfine interaction, being dependent on the position of the nuclear spins, gives rise to many different modulation frequencies. In this work, we obtain the fully relaxed positions of Si and C atoms around the defect from the density functional theory calculation. We then calculate the echo signal from all possible configurations for any number of $^{29}$Si and $^{13}$C within a ~10 Å radius centered on the vacancy. Each configuration is weighted according to their probabilities, determined by the natural abundances of 4.7% and 1.1%, respectively [33]. We treat multiple nuclear spins interacting with a single vacancy independently [34,38], since the nuclear spin-spin interactions are negligible at these timescales. The modulation is largest for nuclear sites far enough away from the vacancy such that the dipole-dipole interaction dominates (*i.e.* beyond NNN) over the Fermi contact, but close enough to have a significant hyperfine interaction.

The theoretical spin echo is plotted with the experimental data in Fig. 4(a-e) with very good agreement, considering the complexity of the nuclear spin environment. The timescales of the modulation match the experiment well, along with the trend of a rapid echo decay at low fields. There is no decoherence included in the theory, so this decay is the result of interfering modulation frequencies that rapidly reduce the echo amplitude [39]. As the magnetic field increases, the echo modulation amplitude decreases as the ratio $A'/\omega_I$ decreases, leading to less initial decay. At these higher fields the nuclear quantization axis approaches the electron spin quantization axis, resulting in no modulation of the electron spin frequency. This behavior does not seem as noticeable for S=1 systems such as the NV center in diamond where nuclear precession in the $m_s$= 0 state is fixed.

The results presented in this Letter provide important advances in understanding the spin properties of the Si vacancy that will be essential to assessing how it may be used for quantum applications. The mapping of the ODMR transitions as a function of magnetic field gives a clearer picture of the different possible transitions and how they can arise in a S=3/2 system. We also present the first ensemble spin echo measurements in this system that show significant echo modulation and a fast decay at low magnetic fields attributed to interference between many nuclear spin configurations with different modulation frequencies. This modulation and decay behavior is reproduced very well by a theoretical model with no fitting parameters that considers all of the possible nuclear spin configurations. At higher magnetic fields, the spin echo decay time is 80 μs, comparable to initial spin echo measurements of NV centers in diamond [5,36] and of single Si vacancies [10], which is a further indication of the utility of the SiC platform for quantum technologies.

This work is supported by the U.S. Office of Naval Research.

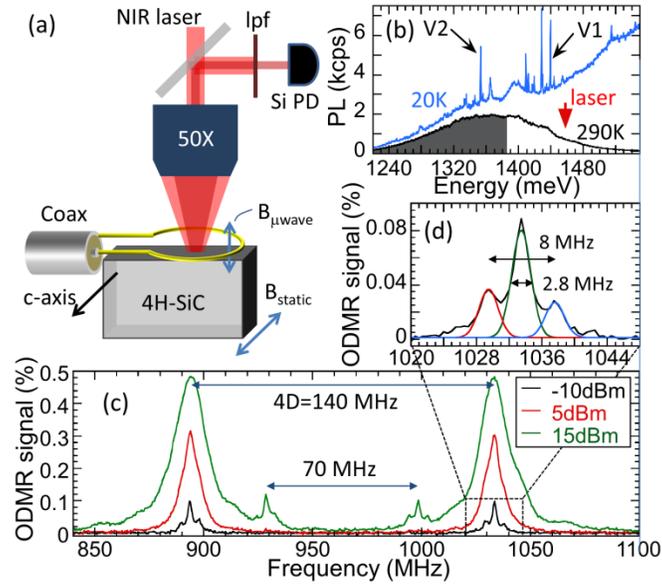

FIG. 1 (color online). (a) Experimental setup for optically-detected magnetic resonance (ODMR). The scattered laser light is filtered from the photoluminescence (PL) by a long pass filter (lpf) and sent to a silicon photodiode (Si PD). (b) PL at 20 K and 290 K, exciting at 532 nm, with 290 K PL scaled down by a factor of 2 for clarity. The typical lpf collection range is shaded in grey, and the typical laser energy is indicated with a vertical arrow. (c) ODMR spectra for B = 34 mT || c  for a series of microwave powers. (d) ODMR at -10 dBm from (c) with the scale expanded to show the hyperfine splitting. Individual Gaussian functions of a three-peak fit are plotted with the spectrum.

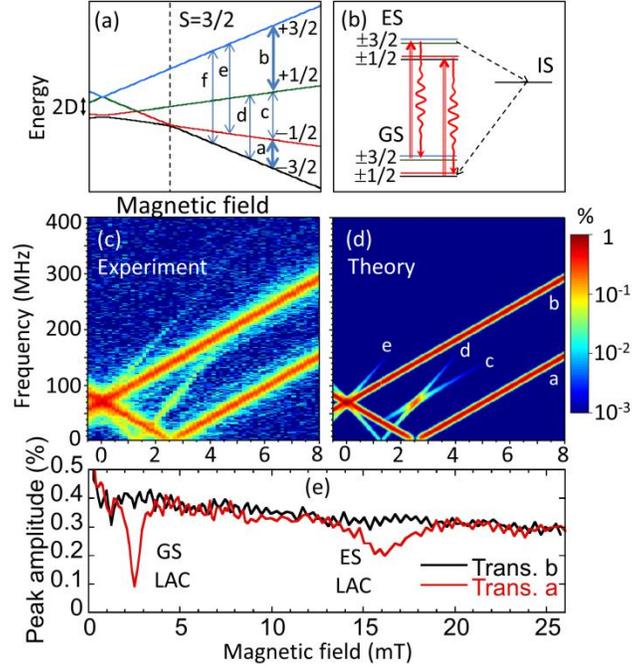

FIG. 2 (color online). (a) Energy levels of the S=3/2 spin states as a function of magnetic field parallel to c. (b) Energy level diagram showing optical excitation and emission between the ground states (GS) and the excited states (ES), along with relaxation to the intermediate states (IS) that result in GS spin polarization. (c) ODMR map as a function of microwave frequency and magnetic field (parallel to c). The color scale is logarithmic. (d) Model calculations of the ODMR with a weak perpendicular magnetic field of 0.05 mT. (e) ODMR amplitude for transitions *a* and *b* as a function of the magnetic field.

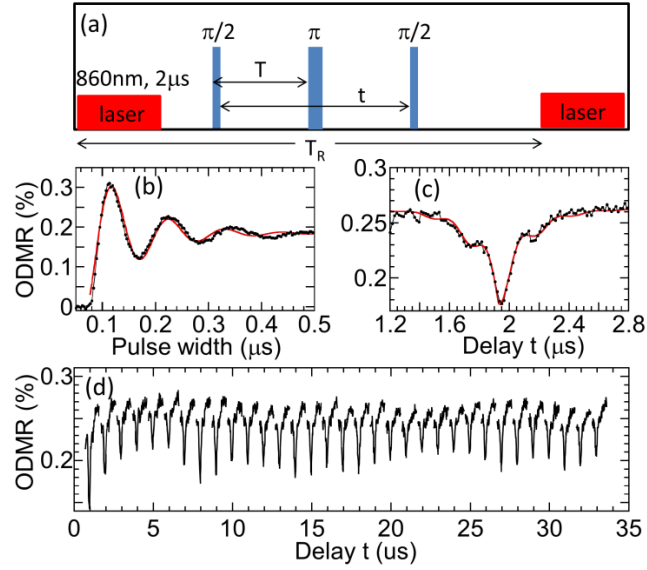

FIG. 3 (color online). (a) Hahn echo sequence with repetition period $T_R$. Narrow blue rectangles represent microwave pulses, and the shorter red rectangles represent laser pulses. (b) Microwave Rabi oscillations with only a single microwave pulse in the sequence, with $T_R = 10$ μs. The red line is a fit to an exponentially decaying cosine (c) Spin echo with $T_R = 40$ μs and $T = 1$ μs. The red line is a fit to $-\exp(-|t - 2T|/T_2^*)$ modulated by a cosine. (d) Spin echo as a function of $t$ for a series of values of $T$, with $T_R = 40$ μs. The echo appears at t = 2T. All measurements are at room temperature with a magnetic field of 34 mT parallel to c and a microwave power of 20 dBm.

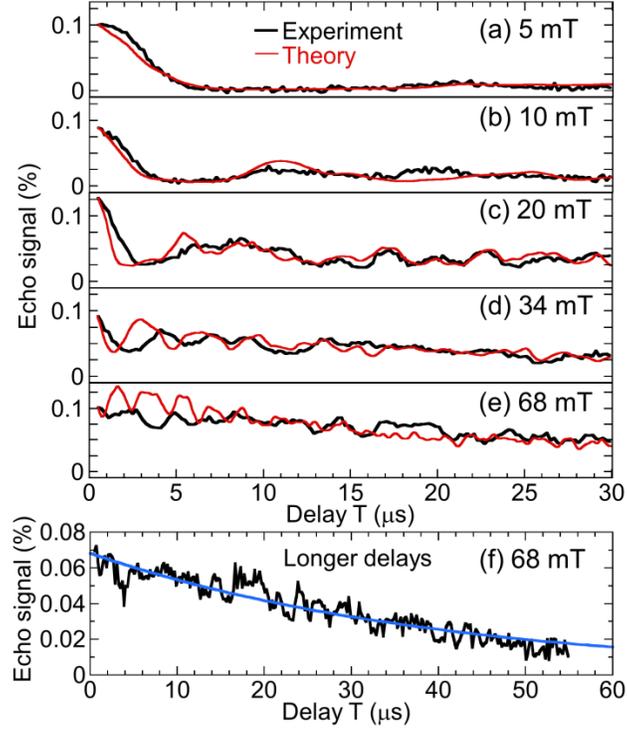

FIG. 4 (color online). (a-e) Experimental and theoretical echo signal as a function of T for a series of magnetic fields parallel to c, with $T_R = 66.7$ μs. The experimental echo is obtained by taking the difference between the signal at $t = 2T + 0.8$ μs and $t = 2T$, and the theory is normalized to best match the experimental data. (f) Echo as a function of T at $B = 68$ mT and $T_R = 125$ μs, with an exponential fit. All measurements are at room temperature.

# Supplemental Material for Spin Coherence and Echo Modulation of the Silicon Vacancy in 4H-SiC at Room Temperature

S. G. Carter[1], Ö. O. Soykal[2], Pratibha Dev[2], Sophia E. Economou[1], E. R. Glaser[1]

[1] Naval Research Laboratory, Washington, DC 20375
[2] National Research Council Research Associate at the Naval Research Laboratory, Washington, DC 20375

**ODMR Model**

The mechanism for ODMR is expected to be similar to that of NV centers in diamond [1] and as described in Refs. [2] and [3]. The ODMR signal is the result of three processes: the spin dependent PL intensity, the optically-induced ground state (GS) spin polarization, and the microwave driving of spin transitions. The spin dependent PL intensity is due to spin selective relaxation from the excited states (ES) to the intermediate states (IS) through an intersystem crossing. This relaxation results in less PL emitted from the ES to the GS for one set of spin states (here, either $m_s = \pm 1/2$ or $\pm 3/2$) [2]. [See Fig. 2(b).] The optically-induced GS spin polarization is the result of relaxation from the IS to the GS that occurs preferentially into particular spin states, polarizing the system into $m_s = \pm 1/2$ or $\pm 3/2$. The microwave magnetic field drives GS spin transitions (as well as ES spin transitions) and tends to equalize the spin populations, resulting in a measured change in PL. ODMR measures the change in PL in the presence of the microwave field. Unlike NV centers in diamond, the PL increases under microwave driving, indicating that optical excitation polarizes the system into the doublet with weaker emission.

To obtain a complete model of ODMR requires a better knowledge of the spin dependent relaxation processes than is currently available, particularly in the present case where field-dependent mixing of GS spin states should alter these processes. Here we take a relatively simple approach to determining how strong particular ODMR lines should be and at what frequency. We first solve the spin Hamiltonian $H = g\mu_B \vec{B} \cdot \vec{S} + DS_z^2 + H_{hf}$ to determine the eigenstates and energies and then make the following assumptions about the strength of the ODMR signal. First, the strength of a particular transition will be proportional to the square of the magnetic dipole between the two states. This assumes that the change in population is linear in the microwave power, ignoring saturation and higher order processes (e.g. two-photon transitions). Second, the effect this change in population will have on the PL is proportional only to the change between the $m_s = \pm 1/2$ and $\pm 3/2$ states. There should be no difference in PL intensity between the $m_s = +1/2$ and $m_s = -1/2$ states or between the $m_s = +3/2$ and $m_s = -3/2$ states. [For this reason, transition *c* ($+1/2 \rightarrow -1/2$) should not appear in ODMR.] However, in the case of mixing between these eigenstates, we assume that it is the difference in population between the *unmixed* eigenstates that gives rise to the change in PL. This assumption is likely an oversimplification, but we expect that it will give qualitative agreement with changes in ODMR intensity due to mixing of eigenstates.

We solve for the eigenstates of the spin Hamiltonian with $g = 2$, $D = 35\,\text{MHz}$, and the Pauli matrices

$$S_x = \frac{1}{2}\begin{pmatrix} 0 & \sqrt{3} & 0 & 0 \\ \sqrt{3} & 0 & 1 & 0 \\ 0 & 1 & 0 & \sqrt{3} \\ 0 & 0 & \sqrt{3} & 0 \end{pmatrix},\quad S_y = \frac{i}{2}\begin{pmatrix} 0 & -\sqrt{3} & 0 & 0 \\ \sqrt{3} & 0 & -1 & 0 \\ 0 & 1 & 0 & -\sqrt{3} \\ 0 & 0 & \sqrt{3} & 0 \end{pmatrix},\quad S_z = \frac{1}{2}\begin{pmatrix} 3 & 0 & 0 & 0 \\ 0 & 1 & 0 & 0 \\ 0 & 0 & -1 & 0 \\ 0 & 0 & 0 & -3 \end{pmatrix}.$$

For a magnetic field parallel to the c-axis ($B = B_z$) and ignoring the hyperfine interaction for the moment, the Zeeman term and $D$ term commute, giving $E_{\pm 1/2} = \pm g\mu_B B_z/2 + D/4$ and $E_{\pm 3/2} = \pm 3g\mu_B B_z/2 + 9D/4$ with spin projections along the z-axis of $m_s = \pm 1/2$ and $\pm 3/2$. For a microwave magnetic field along $y$, the magnetic dipole moment is $\mu_{i,j} = \langle i|S_y|j\rangle$, with non-zero values for transitions with $\Delta m_s = \pm 1$: $\mu_{-3/2,-1/2} = \mu_{+1/2,+3/2} = -i\sqrt{3}/2$ and $\mu_{-1/2,+1/2} = -i/2$.

In the presence of the hyperfine interaction or a stray magnetic field perpendicular to the c-axis, these eigenstates are mixed and other transitions become allowed. For simplicity, we consider the effect of the hyperfine interaction with a next-nearest-neighbor (NNN) $^{29}$Si nuclei, treating it as a classical magnetic field with amplitude $B_{nuc} = A/(2g\mu_B)$. The average nuclear field perpendicular to the c-axis is $\pi/4$ times this value and only 44% of Si vacancies will have at least one NNN $^{29}$Si nuclei. This gives an estimated average nuclear field of 0.05 mT that is included in the calculation of Fig. 2(d) as a field along $y$.

The small perpendicular magnetic field mixes the eigenstates such that additional transitions are allowed and results in a level anticrossing (LAC) between the $m_s = -3/2$ and $m_s = -1/2$ states as shown in Fig. 2(a). We estimate the ODMR signal strength for a transition from state $i$ to state $j$ as

$$ODMR_{i,j} \propto |\mu_{i,j}|^2 |\Delta p_i - \Delta p_j|,$$

where $\Delta p_i = |\langle 3/2|i\rangle|^2 + |\langle -3/2|i\rangle|^2 - |\langle 1/2|i\rangle|^2 - |\langle -1/2|i\rangle|^2$ is the population difference between the unmixed $m_s = \pm 1/2$ and $\pm 3/2$ states. Essentially, the microwave transition must change the relative population of the $m_s = \pm 1/2$ and $\pm 3/2$ states in order to be observed in PL. This also assumes that optical spin polarization always results in the same initial spin polarization. Each transition is given a Gaussian lineshape with FWHM of 6.7 MHz.

As discussed in the main text, the model calculation of ODMR in Fig. 2(d) shows the two main ODMR lines with $\Delta m_s = \pm 1$ as well as the nominally forbidden $\Delta m_s = \pm 2$ lines, which are partially allowed due to mixing from the perpendicular magnetic field. These $\Delta m_s = \pm 2$ lines are both visible near zero field for both the $-3/2 \rightarrow +1/2$ and $-1/2 \rightarrow +3/2$ transitions (*d* and *e*), where there is significant mixing between the $m_s = \pm 1/2$ states. Transition *d* is also visible at higher magnetic fields near the LAC of $m_s = -1/2$ and $m_s = -3/2$ states. At the LAC, the states are strongly mixed, and transitions result in little change in population between $m_s = \pm 1/2$ and $\pm 3/2$, giving a weak ODMR signal for transition *a*. There is good qualitative agreement with experiment except that the appearance of transition *c* at the LAC in the model does not show up well in experiment. This is likely due to the oversimplifications of the model, particularly when the mixing is very strong.

**ODMR dependence on laser excitation energy**

In most of the experiments presented here, the excitation laser was at 850 nm (1459 meV), but there is little change in the ODMR signal at room temperature over a wide range of laser energies. Measurements displayed in Fig. S1 show that the percent ODMR is nearly constant from 1700 meV down to 1475 meV and then increases to a maximum near the V2 ZPL. The total PL intensity is also constant in the higher energy range, but it decreases as the laser approaches the V2 ZPL. This may simply be an indication that PL from other defects is reduced at lower excitation energies, and therefore a higher percentage of the PL comes from V2 under these conditions. No ODMR signal from the V1 vacancy was observed over the entire range of laser excitation energies.

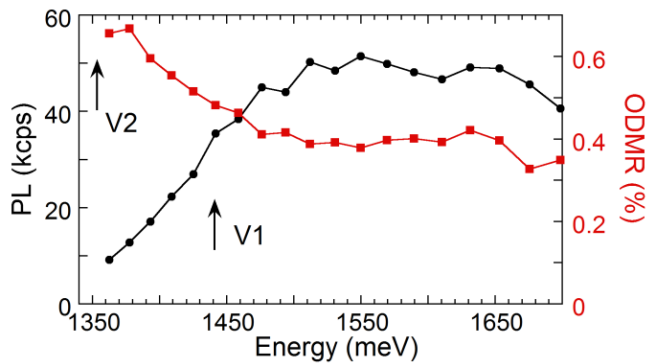

**Figure S1.** Room temperature PL and ODMR signal as a function of the laser excitation energy. Arrows indicate the position of the V1 and V2 ZPL at low temperature. The ODMR was taken at a magnetic field of 28 mT and a constant laser power of 7.5 mW.

**ODMR with B ⊥ c-axis**

An ODMR map for a perpendicular magnetic field is shown in Fig. S2(b), accompanied by the calculated energy levels in S2(a) and the model calculation of the ODMR in S2(c). In this case, the magnetic field strongly mixes the spin states, such that the quantization axis is along the field direction for $g\mu_B B \gg D$. At these higher fields there are two transitions, *a* and *b*, separated by 2D instead of 4D. These transitions are an order of magnitude weaker for an orthogonal field due to the strong mixing of the original spin states. The model reproduces the transition energies well, but the intensities do not agree well with experiment. In particular, lines a and b should be somewhat weaker, and line c should much weaker.

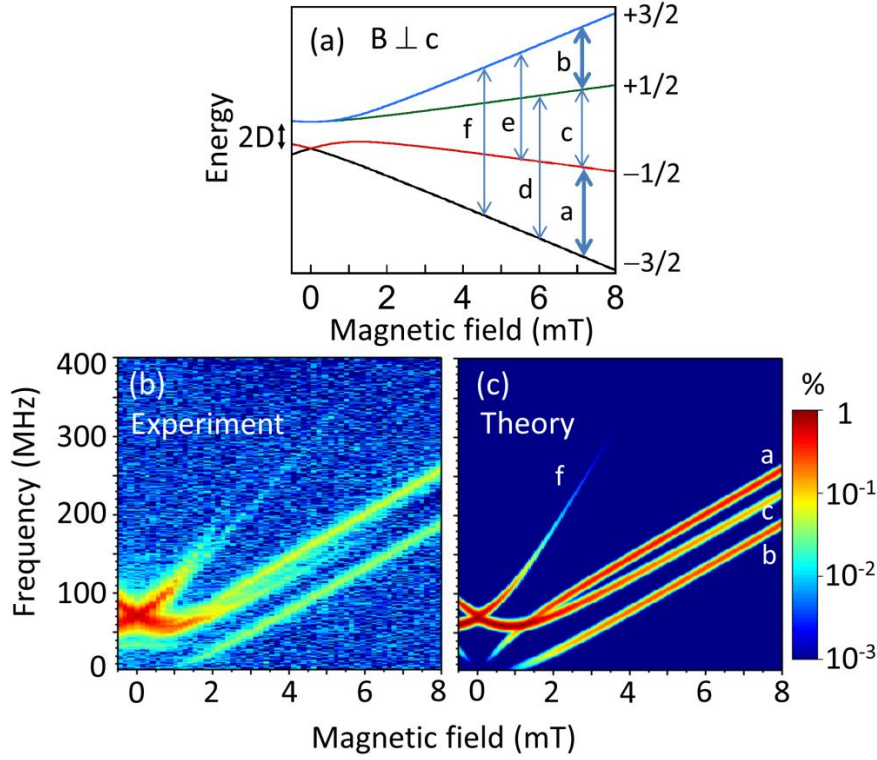

**Figure S2.** (a) Energy levels of the S=3/2 spin states as a function of magnetic field perpendicular to c. The spin states are labeled by the eigenstates for high magnetic field, in which the states are quantized along the magnetic field axis. (b) ODMR map as a function of microwave frequency and magnetic field with the field oriented perpendicular to the c-axis. The color scale is logarithmic. (c) Model calculations of the ODMR.

**Echo Amplitude Modulation: Theory**

The hyperfine interaction between a nuclear spin and the silicon vacancy spin is defined as [4,5],

$$H_{hf} = \mu_0 g_e g_I \mu_B \mu_I \left( \frac{3(I.\hat{r})(S.\hat{r}) - I.S}{4\pi r^3} - \frac{2}{3} I.S |\Psi(0)|^2 \right), \quad (S1)$$

in terms of the electron and nuclear spin magnetic moments. The first term in Eq. S1 is the dipole-dipole interaction between the nuclear and electron magnetic dipole moments. The second term is the Fermi contact interaction, which is due to the non-vanishing electron wave function at the nuclear spin site. The electron Zeeman splitting is much larger than the nuclear Zeeman and hyperfine splitting, which allows us to perform the secular approximation. Under this approximation the vacancy spin state ($m_s=\pm 3/2, \pm 1/2$) along the c-axis cannot be flipped by the nuclear spin. On the other hand, the quantization axis and the energy splitting of the nuclear spin are determined by an effective field consisting of the external magnetic field $B$ and the hyperfine field of the vacancy spin. Under the secular approximation [6], the hyperfine interaction reduces to

$$H_{hf} = [a_f + a_d(3\cos^2\theta - 1)]S_z I_z + a_d(3\sin\theta\cos\theta)S_z I_x = AS_z I_z + A'S_z I_x, \quad (S2)$$

where the dipole-dipole and Fermi-contact hyperfine coefficients are given as $a_d = \mu_0 g_e g_I \mu_B \mu_I/(4\pi r^3)$ and $a_f = 2\mu_0 g_e g_I \mu_B \mu_I |\Psi(0)|^2/3$. The hyperfine interaction is nuclear site-specific depending on the distance $r$ between the vacancy and nuclear spins, and the angle $\theta$ between the c-axis and $r$. The value of the dipole-dipole coefficients in units of MHz are $a_d = 15.720/r^3$ for $^{29}$Si and $a_d = -19.885/r^3$ for $^{13}$C, where $r$ is in Å. The values of $a_f$ are taken from Ref. [8] as 8.7 MHz for NNN $^{29}$Si, 50.0 MHz for the axial position of nearest-neighbor (NN) $^{13}$C, and 44.8 MHz for the basal positions of NN $^{13}$C. Beyond these sites the values of the Fermi-contact coefficients are diminished and are assumed to have negligible effect on echo modulation. In the case of $V_{Si}^-$ ground state spin $S=3/2$, a nearby nuclear spin has four different quantization axes due to the different effective fields created by each state of the vacancy spin. Any change in the vacancy spin state due to the echo pulses causes the nuclear spin to precess around the new effective field, creating an oscillating magnetic field at the vacancy site [7]. After the initial π/2 pulse, interference between the oscillating nuclear fields at the vacancy site will give rise to the amplitude modulation of the echo signal.

Close to the vacancy (e.g. NNN shell for $^{29}$Si), the strong Fermi contact interaction causes the $S_z I_z$ term of the hyperfine interaction to dominate ($A \gg A'$, $\omega_I$) over the $S_z I_x$ term, resulting in diminished echo modulation. Because the Fermi contact interaction drops off quickly as a function of the distance from the vacancy, at intermediate distances the hyperfine dipole term becomes dominant ($a_d \gg a_f$) before eventually becoming negligible at even larger distances (greater than 10Å). Therefore, the modulation effects are strongest at intermediate distances.

The echo signal due to a single nuclear spin species A located at an atomic site $i$ is defined as $X_A[i]$. This site-specific signal is calculated from $X_A[i](T) = \text{Tr}[\hat{\rho}(T).S_z]$, where the density matrix for a given time delay $T$ is $\hat{\rho}(T) = \hat{U}_e^{-1}.\hat{\rho}(0).\hat{U}_e$ in terms of the nutation and free precession operators $\hat{U}_e = \hat{U}_{\pi/2}.\hat{U}_T.\hat{U}_\pi.\hat{U}_T.\hat{U}_{\pi/2}$ of the spin echo pulse sequence. Positions of each atomic site for a relaxed structure are calculated via density functional theory as explained in the next section.

Our system is an ensemble of SiC vacancy centers, and therefore an ensemble average has to be performed in our calculation to account for the different configurations (positions and number of nuclear spins sites per vacancy). The echo signal due to all spatial configurations of a single nuclear spin, involving all the atomic sites $n$ around the vacancy, becomes $X_A^1 = (X_A[1] + X_A[2] + \cdots X_A[n])/P(n,1)$, where the equal probability of each possible configuration is given by the binomial coefficient $P$. Similarly, the echo signal due to multiple nuclear spin possible configurations can be written as $X_A^2 = (X_A[1]X_A[2] + X_A[1]X_A[3] + \cdots + X_A[n]X_A[n'])/P(n',2)$, $X_A^3 = (X_A[1]X_A[2]X_A[3] + X_A[1]X_A[2]X_A[4] + \cdots + X_A[n]X_A[n']X_A[n''])/P(n'',3)$, and $X_A^k = \sum_{n^{(k)}>\cdots>n'>n} X_A[n]X_A[n'] \ldots X_A[n^{(k)}]/P(n^{(k)},k)$ for two, three, etc., nuclear spins, respectively. For a large number of atomic sites involved around the vacancy, each configuration can be well approximated via the mean probability, i.e. $X_A^k = (\sum_{i=1}^n X_A[i]/n)^k$. Therefore, by considering the occupation probabilities of every atomic site by a nuclear spin, the normalized total echo signal due to the nuclear species $A$ becomes

$$X_A^T = \left[\sum_n (N!/(N-n)!n!)q^n(1-q)^{N-n}X_A^n\right] / \left[\sum_n (N!/(N-n)!n!)q^n(1-q)^{N-n}\right]$$

in terms of the natural abundance ratio $q$ and the total number of atomic sites $N$ considered. In the case of $V_{Si}^-$ in 4H-SiC, both $^{13}$C and $^{29}$Si are assumed to simultaneously interact with the vacancy spin and give the signal modulation determined by $X_C^T X_{Si}^T$. Higher order nuclear spin-spin interactions are ignored in the signal calculations due to their much longer timescales.

**Density functional theory calculations**

The echo signal was calculated using nuclear positions obtained from the spin-polarized calculations based on density-functional theory (DFT). In order to account for the exchange-correlation effects, we used the generalized gradient approximation [9] of Perdew-Burke-Ernzerhof (PBE) [10]. These calculations were performed using the Quantum-ESPRESSO package [11]. We used a kinetic energy cutoff of 40 Ry for expanding the wavefunctions and a cut-off of 350 Ry for charge densities. The Monkhorst-Pack scheme [12] was used to generate the Γ-centered, 2X2X2 k-point grid for a 6X6X2 supercell (576 atoms) of 4H-SiC.

In order to obtain the nuclear positions used in this work, the structure was relaxed after creating a charged silicon vacancy, with the relaxation threshold set to be better than $10^{-4}$ Ry/a.u..